\documentclass[utf8,12pt]{article}
\usepackage{sbc-template}
\usepackage{graphicx,url}
\usepackage[utf8]{inputenc}
\usepackage[english]{babel}
\usepackage{comment}
\usepackage{booktabs}
\usepackage{tabularx}
\usepackage{fancyvrb}
\usepackage{tcolorbox}
\usepackage{longtable}
\usepackage{array}
     
\title{A Systematic Security Testing Approach for\\ InterUSS-based environments}

\author{Henrique Curi de Miranda\inst{1}\inst{2}, Ágney Lopes Roth Ferraz\inst{1}, \\ Wagner Comin Sonaglio\inst{1} and Lourenço Alves Pereira Júnior\inst{1}}

\address{Divisão de Ciência da Computação -- Instituto Tecnológico de Aeronáutica
  (ITA)\\
  Praça Marechal Eduardo Gomes, 50 -- 12228-900 -- São José dos Campos -- SP -- Brasil
\nextinstitute
  Tempest Security Intelligence\\
  Rua da Alfândega, 35 -- Loja 216A -- 1º Piso -- Recife -- PE -- Brasil
  \email{henriquecuri1@proton.me, roth@ita.br, sonaglio@ita.br and ljr@ita.br}
}

\begin{document} 

\maketitle

\begin{abstract}
Unmanned Traffic Management (UTM) federated ecosystems, such as InterUSS, enable secure coordination among UAS Service Suppliers (USSs). However, they bring up some security challenges at the infrastructure level that haven’t been fully explored. This paper presents a security testing approach for InterUSS-based environments from the maintainer's perspective. By deploying and analyzing a working InterUSS infrastructure, we pinpoint key components and develop specific security tests aligned with established standards and protocols, such as mTLS and OAuth 2.0. We compiled these tests into a Testing Guide that aids both in validating components and in analyzing interactions across InterUSS-based ecosystems, filling a gap in current research.
\end{abstract}

\section{Introduction}

In recent years, the rapid growth of Unmanned Aerial Vehicles (UAVs) has elevated traffic management to a primary concern. Among Unmanned Traffic Management (UTM) solutions, the InterUSS project, maintained by the Linux Foundation, stands out as a federated platform that enables secure and efficient information exchange through its Discovery and Synchronization Service (DSS). This architecture represents a significant advancement over traditional aircraft management technologies. Nonetheless, novel security challenges emerge with decentralized architectures \cite{rv}.

The project implementation targets compliance with \cite{3548} and \cite{3411}, and also implements the European concept of U-Space for UTM. While U-Space defines the regulatory and operational environment under \cite{2021664}, InterUSS provides the necessary digital infrastructure to enable interoperability among multiple service providers. However, a complete UTM ecosystem relies on integrating several other critical components that fall under the responsibility of environment administrators. Consequently, each ecosystem possesses unique particularities that, if poorly implemented, may introduce severe security vulnerabilities. 

While security testing is recognized as one of the most effective ways to prevent malfunctions in UAV cybersecurity, current literature provides only a partial and vague description of testing processes. As noted in the recent systematic review \cite{literature_review}, existing studies often address a few specific particularities, leaving substantial room for improvement in practical implementation. To address these challenges, this paper proposes a specific methodology tailored to InterUSS-based environments to support the evaluation and maintenance of a security-oriented infrastructure. Unlike existing general frameworks, our approach focuses on the perspective of the ecosystem’s maintaining authority, providing a comprehensive survey of critical cybersecurity components. Although multiple frameworks exist, none are specifically designed for this kind of environment.

In summary, our work contributes to the field by providing a structured path for ecosystem security, ensuring that federated architectures remain resilient against emerging threats. The remainder of this paper is organized as follows: Section \ref{sec:2} presents related work. Section \ref{sec:3} describes the testing environment and the security of InterUSS critical components. Section \ref{sec:4} discusses the proposed testing approach.  Finally, Section \ref{sec:5} presents the conclusions and future works.

\section{Related Work}
\label{sec:2}

\textbf{UAV and UTM Security Surveys}: Several studies have focused on documenting the vulnerability landscape of unmanned aerial vehicles (UAVs) and unmanned traffic management (UTM) systems. Works such as \cite{16} and \cite{9} provide comprehensive analyses of vulnerabilities and mitigation strategies, mainly focused on Unmanned Aerial Vehicle (UAVs), with some also exploring UTM-related interactions. Similarly, \cite{2} focuses specifically on the distinction between commercial and military drones. In the same vein, \cite{15} and \cite{13} examine drone-related architectures and related systems; specifically, the latter describes the European project COMP4DRONES and discusses security concerns. Furthermore, \cite{1} emphasizes the critical need for risk assessment within UTM systems and contributes with a model for risk assessment. Despite their breadth, these surveys often offer a generalist view or focus primarily on the aerial vehicles themselves. In contrast, our work addresses the unique architectural complexities of the InterUSS federated ecosystem, shifting the focus to the infrastructure-level security required by environment administrators, as shown in Table \ref{tab:works}.

\textbf{Offensive Testing and Threat Modeling}: Another significant research line explores security from an offensive perspective. Studies such as \cite{7}, \cite{10}, and \cite{11} present practical security testing results against specific drone models, while \cite{4} adopts a broader offensive approach. Regarding threat identification, \cite{3} and \cite{6} propose modeling techniques to explore emerging trends and issues. Nevertheless, these methodologies are frequently model-specific or focused on individual vehicle exploits. Our approach circumvents these limitations by proposing a methodology tailored to the InterUSS environment, enabling a broader assessment of the ecosystem’s resilience rather than isolated components.

\textbf{Detection and Methodology Frameworks}: Efforts to improve detection techniques or security evaluation have led to the development of various frameworks. For instance, \cite{12} and \cite{17} introduce detection and mitigation frameworks for drone-based attacks. In the field of testing methodologies, \cite{8} applies Purple Teaming and proposes a kill chain model, while \cite{5} introduces a customizable tool for drone security testing. \cite{14} proposes ARES, a vulnerability assessment framework for Robotic Aerial Vehicles, which also includes UAVs. However, these proposals are often designed in a generic manner, lack focus on infrastructure testing, or lack specific alignment with the federated nature of modern UTM platforms. To address this challenge, we propose a specific testing methodology from the perspective of the maintaining authority, focusing on the critical intersection between architecture and security infrastructure.

In contrast to previous efforts, we propose a new security testing framework designed specifically for InterUSS-based environments. This framework enables the validation of individual components and their dynamic behavior within the environment, from the perspective of the ecosystem's maintaining authority. In this work, we present a methodology that enables a more effective and personalized security testing approach for this type of ecosystem, thereby improving existing approaches.

\begin{table}[t]
\caption{Related works comparison}
\label{tab:works}
\centering
\resizebox{\textwidth}{!}{
\begin{tabular}{@{}lcccccc@{}}
\toprule
          & \multicolumn{1}{l}{\textbf{Security issues}} & \multicolumn{1}{l}{\textbf{Mitigations}} & \multicolumn{1}{l}{\textbf{Security testing techniques}} & \multicolumn{1}{l}{\textbf{Testing methodology}} & \multicolumn{1}{l}{\textbf{UTM related}} & \multicolumn{1}{l}{\textbf{InterUSS or U-Space}} \\ \midrule
\cite{1}         & X                                            & X                                        &                                                          &                                                  & X                                        &                                                  \\
\cite{2}         & X                                            & X                                        & X                                                        &                                                  &                                          &                                                  \\
\cite{3}         & X                                            &                                          & X                                                        &                                                  &                                          &                                                  \\
\cite{4}         & X                                            &                                          & X                                                        &                                                  &                                          &                                                  \\
\cite{5}         & X                                            &                                          & X                                                        &                                                  &                                          &                                                  \\
\cite{6}         & X                                            & X                                        &                                                          &                                                  &                                          &                                                  \\
\cite{7}         & X                                            &                                          & X                                                        &                                                  &                                          &                                                  \\
\cite{8}         & X                                            &                                          & X                                                        & X                                                &                                          &                                                  \\
\cite{9}         & X                                            & X                                        & X                                                        &                                                  &                                          &                                                  \\
\cite{10}        & X                                            & X                                        & X                                                        &                                                  &                                          &                                                  \\
\cite{11}        & X                                            & X                                        & X                                                        &                                                  &                                          &                                                  \\
\cite{12}        & X                                            &                                          &                                                          &                                                  &                                          &                                                  \\
\cite{13}        & X                                            & X                                        &                                                          &                                                  & X                                        & X                                                \\
\cite{14}        & X                                            &                                          & X                                                        & X                                                &                                          &                                                  \\
\cite{15}        & X                                            & X                                        &                                                          &                                                  &                                          &                                                  \\
\cite{16}        & X                                            & X                                        & X                                                        &                                                  & X                                        &                                                  \\
\cite{17}        & X                                            & X                                        & X                                                        &                                                  & X                                        &                                                  \\\midrule
This work & X                                            & X                                        & X                                                        & X                                                & X                                        & X                                                \\
\bottomrule
\end{tabular}
}
\end{table}

\section{Testing environment and Critical components}
\label{sec:3}

To conduct the tests and develop a Security Testing Approach, the first step was to implement an InterUSS-based environment with all necessary components. The environment also needed to be fully functional, allowing for service interaction and observation of its operational dynamics.

For testing purposes, \cite{monitoring} was utilized. This is an official InterUSS project repository that provides several testing tools, including the capability to create a functional environment. To initialize the ecosystem's core services, the Makefile in the root directory was executed using the following command:

\begin{tcolorbox}[center, halign=center,width=\textwidth]
{\tt sudo make start-locally}
\end{tcolorbox}

As a result, a series of checks is performed, and the 'build/dev/run\_locally.sh' script is executed with the 'up' argument. Executing this script invokes the Docker Compose file located at 'build/dev/docker-compose.yaml', which uses Docker to create the necessary networks and deploy the service containers locally.

At this stage, the environment remains passive, awaiting requests, as it does not yet have USSs to interact with. The same repository provides several USS mocks for testing purposes, which are also initialized via the Makefile in the root directory. The command used is as follows:

\begin{tcolorbox}[center, halign=center,width=\textwidth]
{\tt sudo make start-uss-mocks}
\end{tcolorbox}

Similar to the previous workflow, with just one additional step, the Makefile calls the 'monitoring/mock\_uss/start\_all\_local\_mocks.sh' file. This, in turn, calls 'monitoring/mock\_uss/run\_locally.sh', which finally executes 'monitoring/mock\_uss/docker-compose-yaml' to bring the containers up. After these steps, a functional InterUSS environment will be operational.

\subsection{Critical components}

Based on an analysis of the complete environment's architecture and operational dynamics, it is possible to identify individual components and subsystems that require greater security attention.

From a networking perspective, the ecosystem relies on resources that must be publicly exposed to function properly, enabling service discovery and interaction. However, there are internal services whose public exposure constitutes a security risk and could facilitate malicious actors' actions. In this case, careful attention must be given to proper component segregation, restricting access to certain resources exclusively to administrators and previously authorized essential services. It is imperative for the ecosystem that a robust network architecture be implemented, utilizing a multi-layered design based on Zero Trust principles, as highlighted in \cite{zero_trust}. This approach enables the segmentation of services into distinct visibility zones, thereby mitigating the actions of malicious actors.

The core of the DSS service consists of the database, typically CockroachDB, which enables data persistence and maintains a representation of the airspace. The InterUSS ecosystem operates under a federated structure, where authority and responsibility are shared, thereby eliminating single points of failure. Consequently, the implementation follows an integrated database cluster model, providing redundancy and geographic distribution, which contributes to high availability and low latency. Data synchronization among redundant cluster nodes is achieved via the Raft consensus algorithm through a MultiRaft architecture, which optimizes large-scale replication. According to \cite{raft}, this algorithm enables management of replicated logs in a distributed system by using leader-election logic that is reinitiated upon failure. In the InterUSS context, databases communicate with one another using the gRPC protocol over HTTP/2 for message exchange between nodes; thus, ensuring the confidentiality and integrity of these messages is vital to prevent malicious actors from manipulating airspace representation data. Furthermore, USSs must be able to interact with the database to insert operational information and discover data from other USSs. It is essential to adopt techniques that prevent malicious database interactions, particularly those related to injection vulnerabilities, and to use secure methods to protect data at rest. It is worth noting that interactions do not occur directly with the database but rather through gateways that mediate communication without exposing the cluster.

The operational security of the InterUSS ecosystem relies on the authentication and authorization of the entities interacting within the environment. The OAuth 2.0 framework functions by delegating authorization, allowing an external application (typically a USS) to obtain limited access to services. The OAuth authorization server is responsible for verifying the requester's identity, which typically undergoes a certification and onboarding process before joining the ecosystem. The protocol defines different flows for obtaining tokens (OAuth Grant Types); the Client Credentials Grant is typically utilized in an OAuth server integrated into the InterUSS environment, as data exchange normally occurs automatically and exclusively via API. It is important to note that, since the authority responsible for the ecosystem may implement modifications that best suit its specific use case, other Grant Types might be employed, which could create insecure flows. The paper by \cite{oauth} reinforces the idea that security issues with the OAuth 2.0 protocol primarily stem from implementation choices that depend on the environment's particularities. The OAuth server is also responsible for implementing different access profiles through the limits defined by the generated token, referred to as the 'scope.' In this sense, it is of paramount importance to ensure that a given USS operates only with the authorized permission level, thereby avoiding excessive privilege.

Upon authentication with the OAuth server, USSs receive JSON Web Tokens (JWT) that enable stateless session control within the ecosystem. The server precludes the need to maintain active user lists by utilizing cryptographic signature validation to attest to the token's integrity and authenticity. Therefore, it is imperative to employ robust signing algorithms and ensure that the private key is maintained in a secure environment. Beyond the signature, the service must mandatorily validate the 'scope' and 'audience' claims. To mitigate risks in the event of a token leakage, tokens must have a short expiration time, preventing malicious actors from interacting with the InterUSS environment for prolonged periods. Finally, if USS interaction with the ecosystem can also occur via a web application, the browser storage strategy must be rigorously defined to prevent security issues.

Within the InterUSS ecosystem, most web applications are USS portals, which, as they are not maintained by the authority responsible for the environment, fall outside the scope of this work. However, the authority likely maintains its own web applications, whose purposes depend on its specific requirements, for example, public data disclosure or the registration of interested operators. Additionally, web applications for service management, such as databases, are common and, as previously discussed from a networking perspective, require access restrictions. In particular scenarios, interaction with the environment may be permitted not only via API but also through web applications, even though this is not the standard expected by the InterUSS project. Web applications frequently prove fertile ground for security issues; as argued in \cite{hackekshandbook}, applications of this nature are often developed, in whole or in part, with custom code, making them highly diverse and, consequently, prone to a variety of security flaws. Typical vulnerabilities in these applications are extensively studied and include, but are not limited to: Cross-Site Scripting (XSS), Server-Side Request Forgery (SSRF), and Broken Access Control. Due to their significance and unique peculiarities, numerous security testing frameworks have been developed and widely validated, proving effective across different scenarios.

In a complex ecosystem characterized by various actors in constant interaction, tracking behaviors is not merely a best practice but an essential security requirement. To this end, the logging policy constitutes the foundation of monitoring, necessitating careful planning: data must be granular enough to identify operational failures and malicious actions with agility. Additionally, the robustness of this policy depends on secure storage, where access is restricted to authorized profiles. To ensure full integrity, logs must be immutable and adhere to the WORM (Write-Once, Read-Many) principle. This ensures that, once recorded, data cannot be altered or deleted, thereby serving as a reliable audit trail.

\begin{figure}[htpb]
\centering
        \includegraphics[width=\linewidth]{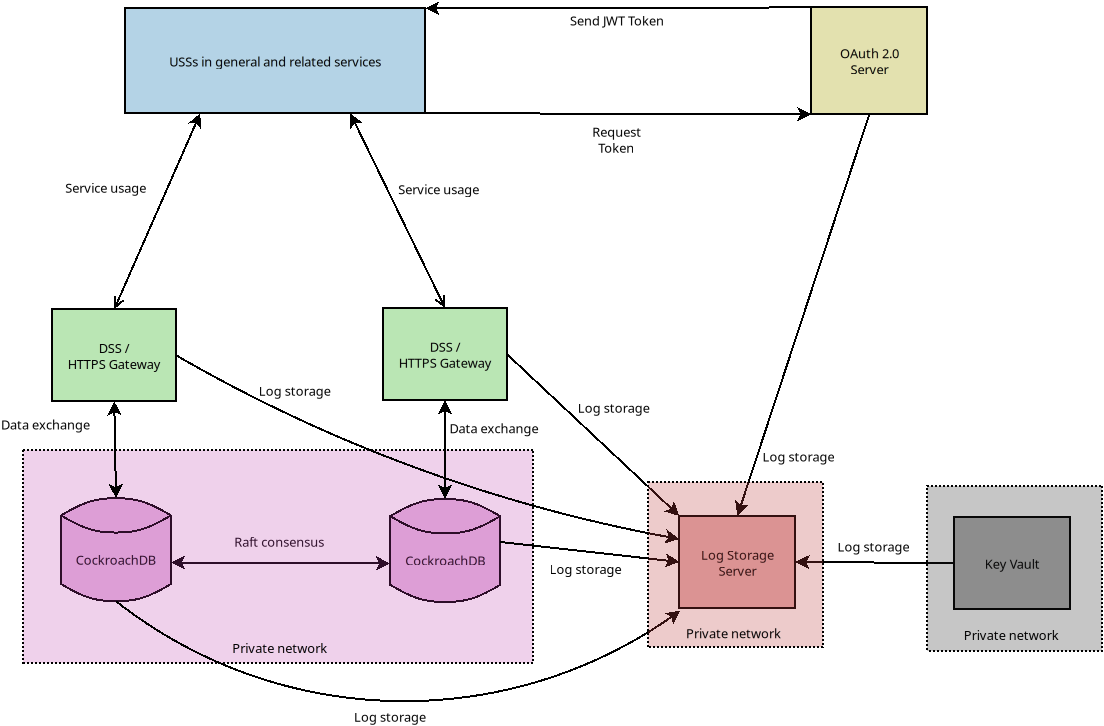}
\caption{Typical InterUSS architecture} 
    \label{fig:diagram}
\end{figure}

UAS Service Suppliers (USSs) are the actors that must interact with the ecosystem in various ways, discovering one another and exchanging information. Nevertheless, their internal components and specificities fall outside the scope of InterUSS ecosystem administrators and, therefore, are not considered part of this work. However, the manner in which USSs interact with environmental components and the interfaces they use are within the scope of this study, including scenarios in which these interactions occur solely via APIs or through web applications. Table \ref{tab:critical_components} summarizes the identified critical components and main security concerns. Figure \ref{fig:diagram} summarizes a typical InterUSS environment and the interactions among its components.

\begin{table}[ht!]
    \centering
     \caption{Critical components and main concerns}
    \begin{tabular}{|c|c|} \hline
       \textbf{Component} & \textbf{Concerns}\\ \hline
       & \\
       \textbf{Network} & - Components segregation\\ 
       & \\ \hline
       & \\
       & - Confidentiality on message exchange\\
       & - Integrity on message exchange\\
       \textbf{Database Cluster} & - Avoid injection vulnerabilities\\
       & - Protect data at rest\\ 
       & \\ \hline
       & \\
       & - Secure authorization flow\\
       \textbf{OAuth Server} & - Correct permission level for profiles\\ 
       & \\ \hline
       & \\
       & - Robust signing algorithm\\
       & - Private key properly stored\\
       & - Signature validation\\
       \textbf{JWT Token}& - Scope validation\\
       & - Audience validation\\
       & - Short expiration\\
       & - Storage on browser in case of web interface\\ 
       & \\ \hline
       & \\
       & - Avoid administrative interface exposure\\
       \textbf{Web Applications} & - Typical web vulnerabilities\\ 
       & \\ \hline
       & \\
        & - Data must be granular\\
       \textbf{Logs} & - Secure storage\\
        & - Must be immutable\\ 
        & \\ \hline
    \end{tabular}
    \label{tab:critical_components}
\end{table}

\section{Testing Approach}
\label{sec:4}

The identification of critical components and their systemic organization enables both individual and integrated analyses of environmental interactions. The investigation of vulnerabilities, coupled with the proposal of mitigation and best-practice measures, underpins a rigorous testing methodology. This approach validates each component individually while accounting for systemic interdependencies, thereby ensuring the depth and effectiveness of the implementation.

\subsection{Network testing}
\label{subsec:networktesting}

Network validation must ensure that critical components remain externally inaccessible: log repositories, administrative interfaces, key management services, and database nodes. In the case of the database cluster, the proper functioning of the InterUSS environment requires constant communication between nodes to exchange information. Ideally, these should be maintained on an isolated network with restricted communication; however, the project's federated nature often makes this model infeasible. In such cases, node access must be restricted to specific, strictly necessary addresses for the environment's operation through an IP Allowlisting policy. On the other hand, components such as the OAuth authorization server, HTTPS Gateways, and public-facing web applications require exposure for interoperability and do not inherently compromise security.

Table \ref{tab:network} summarizes the components that must be restricted and those that should be publicly exposed. It is essential to note that different implementations may introduce new components, which must undergo an individualized risk analysis based on their attack surface and operational criticality to determine their appropriate placement within the proposed segmentation.

\begin{table}[t]
    \centering
    \caption{Network visibility zones}
    \begin{tabular}{|c|c|} \hline
        & \\
         & - OAuth Authorization Server\\
         \textbf{Public Access} & - HTTPS Gateways\\
         & - Public-facing Web Applications\\ 
         & \\ \hline
         & \\
         & - Log Repositories\\
         \textbf{Restrict Access} & - Key Management Services\\
         & - Database Nodes\\ 
         & \\ \hline
    \end{tabular}
    \label{tab:network}
\end{table}

\subsection{Database cluster testing}

The adoption of a decentralized architecture with redundant database nodes necessitates continuous synchronization via consensus and discovery protocols. In this scenario, it is imperative that all participating nodes be previously identified and authenticated. To this end, the Mutual Transport Layer Security (mTLS) protocol is used, ensuring mutual entity identification and guaranteeing the confidentiality and integrity of data in transit via TLS 1.3. A rigorous validation of the implementation's compliance with this protocol for inter-node communication is recommended.

As described in subsection \ref{subsec:networktesting}, database nodes must be isolated from public access and operate within private networks or be protected by Access Control Lists (IP Allowlisting). However, the ecosystem’s functionality requires USSs to interact with stored data; for this purpose, HTTPS Gateways are employed. SQL injection threats, as described in \cite{hacking_api}, are a classic security vulnerability and must be tested to prevent external actors from interacting arbitrarily through the HTTPS Gateway.

In addition, data-at-rest protection is recommended and is natively supported by CockroachDB. The implementation of this feature must be tested with a secure algorithm. The use of the AES-256 algorithm is recommended to ensure compliance with high-criticality cybersecurity standards.

\subsection{OAuth Server testing}

In the InterUSS ecosystem, most interactions consist of Machine-to-Machine (M2M) communications; consequently, the most widely used authorization flow is the Client Credentials Grant. In the proposed architecture, the typical relationship is established between a previously identified and authorized Service Provider (USS) and the responsible authority.

Although the Client Credentials flow can use a \textit{client\_secret}, the preferred practice is to replace it with asymmetric-key-based authentication (e.g., mTLS), as formally defined in \cite{rfc_oauth_mtls}. In this scenario, the USS is responsible for generating and maintaining a public/private key pair, where the private key remains protected while the public key is shared with and stored by the authorization server. Authentication occurs via a Mutual Authentication Handshake, followed by a request for a JWT that becomes a 'Sender-Constrained Token'. The server must be tested to ensure correct mTLS implementation.

In ecosystems with a large number of participants, certificate maintenance can be challenging, and the responsible authority may opt to use a \textit{client\_id} and \textit{client\_secret}. In this case, the \textit{client\_secret} functions as an application password; therefore, its generation process and storage by the authorization server must be tested. The secret must be generated using a cryptographically secure algorithm with at least 256 bits of entropy, and it must be stored using a secure hashing algorithm, never in plaintext.

If other authorization flows are implemented, they must also be tested individually to detect insecure or non-recommended implementations. Such a decision may be motivated by adopting a web interface for interacting with the ecosystem, in which case specific tests for this scenario (such as testing for Cross-Site Request Forgery (CSRF)) become necessary. The OAuth server must also implement the Principle of Least Privilege (PoLP) for access profiles and should be tested for profiles with excessive permission scopes.

\subsection{JWT Token testing}

Once obtained, the token can be decoded from Base64URL, allowing for the identification and visualization of its three components: header, payload, and signature. For subsequent tests, the JWT can be modified by decoding, modifying, and re-encoding.

The first security test consists of verifying its expiration time. The 'exp' claim in the payload contains a value, in Unix timestamp format, representing the token's exact validity, as specified in \cite{rfc_jwt}. Ideally, JWTs should be short-lived, typically between 5 and 15 minutes, thereby forcing constant identity revalidation, however, the use of a "SenderConstrained Token" mitigates this problem. After verifying the validity as declared in the payload, it is also necessary to identify the services that use the token and check whether expired tokens are indeed being rejected or continue to be accepted.

Since the use of JWTs precludes the need for active session storage (stateless), token validation and integrity follow signature validation logic. In this sense, several security flaws can be exploited and must be targeted for testing. The signature algorithm must be robust; therefore, it is necessary to verify that it is considered secure (e.g., RS256).

Even if a secure algorithm is used, improper exposure of the signing private key completely compromises its purpose. Thus, it is essential to understand how the key is stored and ensure it has not been compromised, complementing the previously described network tests on \ref{subsec:networktesting}. It is necessary to test services for signature validation, analyzing whether omissions or arbitrary alterations cause a token to be accepted by the application.

'None algorithm' tests must be conducted by changing the algorithm declared in the header to 'None' to verify if services fail by ignoring signature validation. Another possible vulnerability occurs when an attacker can force signature verification with a different algorithm, leading to the attack known as 'Algorithm Confusion.' This typically involves switching from an asymmetric algorithm to a symmetric one, leveraging the public key as an HMAC secret key for signature verification, thereby allowing the attacker to forge valid tokens.

Beyond token legitimacy and validity, services must be tested for 'scope' and 'audience' (aud) verification, ensuring that the bearer performs only actions belonging to its profile within the respective applications, thereby preventing privilege escalation attacks. This is a critical point for a federated architecture, such as InterUSS, as it prevents tokens from being reused across different services.

If the authority responsible for the environment decides to permit access via a web interface, generated tokens must be stored in session storage or as a secure cookie (Secure, HttpOnly, SameSite=Strict) and never in local storage; therefore, this testing must also be conducted. It is important to emphasize that in a typical environment where interactions occur exclusively via API, this test becomes no longer applicable.

\subsection{Web Applications testing}

Due to the high complexity and unique particularities of each web application, it is essential that they be addressed and tested individually. Since they constitute a large portion of existing applications and also harbor a significant share of vulnerabilities, security testing methodologies for them tend to be complex. Applications of this nature are not a novelty, nor are testing frameworks, with many having been proposed and extensively validated over the years.

Taking advantage of this fact, testing web applications within the InterUSS ecosystem should involve selecting a well-known methodology that best fits the environment's specific needs. As a suggestion, \cite{wstg} is one of the most widely recognized and used guides for this purpose, covering many vulnerabilities that must be verified in applications integrated into the InterUSS ecosystem.

It is essential to emphasize that any testing must be conducted through human analysis, utilizing available tools as aids rather than as the final authority; fully automated analyses are insufficient for this purpose. In the same vein, while the methodology presents a series of structured tests, other relevant, non-encompassed tests relevant to the environment must also be performed. Furthermore, the use of any testing framework must involve a degree of adaptation to the application environment, focusing on aspects that best align with the technologies and techniques employed in application construction.

\subsection{Log policy testing}

In the InterUSS environment, proper log management is vital for monitoring and auditing both cyberspace and the physical airspace, recording, for instance, interaction data between USSs and infrastructure, communication among backend services, and external interactions through gateway services. Standards such as \cite{3548} define strict requirements for recorded data; therefore, testing for compliance with these standards is mandatory.

In conjunction with tests involving the JWT and the OAuth server, integrity and non-repudiation must be verified to ensure accurate identification of the agents involved in a generated record and to prevent interception and modification. The use of WORM (Write-Once, Read-Many) storage systems is imperative and must be tested. Techniques such as Hash Chaining provide a robust tool for preventing and easily detecting tampering attempts. In addition to network testing on \ref{subsec:networktesting}, the absence of external access vectors to stored logs must be verified, ensuring access is restricted to necessary profiles through segregated means.

\subsection{Discussion}

The identification of critical components, potential security threats, and pertinent tests constitutes a Testing Approach methodology, outlined in a Testing Guide and summarized in Table \ref{tab:methodology}. This Testing Approach offers a clear roadmap for identifying and mitigating issues within the environment. It also addresses unique challenges common in projects like InterUSS. Like any methodology, it can be expanded or adjusted to suit the specific needs of the ecosystem being tested.

\begin{longtable}{|p{0.22\linewidth}|p{0.72\linewidth}|}
    \caption{Critical components and security tests}
    \label{tab:methodology} \\
    \hline
    \textbf{Component} & \textbf{Security tests} \\
    \hline
    \endfirsthead
    
    \caption[]{Critical components and security tests (continued)} \\
    \hline
    \textbf{Component} & \textbf{Security tests} \\
    \hline
    \endhead
    
    \hline
    \multicolumn{2}{|r|}{Continued on next page} \\
    \endfoot
    
    \hline
    \endlastfoot
    
    \textbf{Network} & 
    \begin{minipage}[t]{\linewidth}
    \begin{itemize}
        \item Check if restricted access services are in a private network or with access through IP Allow-list
        \item Check if public access services are in a public network
    \end{itemize}
    \end{minipage} \\
    \hline
    
    \textbf{Database Cluster} & 
    \begin{minipage}[t]{\linewidth}
    \begin{itemize}
        \item Test mTLS implementation
        \item Test for SQL Injection on HTTPS Gateways
        \item Test for data-at-rest protection
        \item Test encryption algorithm security
    \end{itemize}
    \end{minipage} \\
    \hline
    
    \textbf{OAuth Server} & 
    \begin{minipage}[t]{\linewidth}
    \begin{itemize}
        \item Test mTLS implementation
        \item Check profiles' permission scopes according to PoLP
        \item If client\_secret, test entropy
        \item If client\_secret, check storage using hash algorithm
        \item If other authorization flows, test for insecure ones
        \item If web interface, test for web vulnerabilities
    \end{itemize}
    \end{minipage} \\
    \hline
    
    \textbf{JWT Token} & 
    \begin{minipage}[t]{\linewidth}
    \begin{itemize}
        \item Test for long expiration time token
        \item Check if expired tokens are not being accepted
        \item Test signature algorithm
        \item Search for private key exposures
        \item Check signature validation
        \item Test for 'None algorithm' attack
        \item Test for 'Algorithm Confusion'
        \item Check scope validation
        \item Check audience validation
        \item If web interface, test for improper token storage
    \end{itemize}
    \end{minipage} \\
    \hline
    
    \textbf{Web Applications} & 
    \begin{minipage}[t]{\linewidth}
    \begin{itemize}
        \item Manually and individually test each one through a well-known methodology
    \end{itemize}
    \end{minipage} \\
    \hline
    
    \textbf{Logs} & 
    \begin{minipage}[t]{\linewidth}
    \begin{itemize}
        \item Check log granularity through compliance requirements
        \item Check WORM policy
        \item Test for malicious data corruption
        \item Test for improper external access
    \end{itemize}
    \end{minipage} \\
    \hline
\end{longtable}

As a proof of concept, the deployed InterUSS testing ecosystem was used to evaluate the methodology, despite its intentionally insecure design. It was possible to identify some security problems, such as: insecure communication between database nodes, improper OAuth 2.0 authorization flow, and private key exposure. 

\section{Conclusion and future works}
\label{sec:5}

This work presented a testing methodology for InterUSS-based environments as a testing guide. The proposed methodology identifies critical components within the InterUSS ecosystem, their mutual interactions, and interactions with external actors. It also identifies security issues, corresponding diagnostic tests, mitigations, and best practices. The presented Testing Approach broadly encompasses the inherent complexities of an InterUSS-based environment, analyzing each component individually and the system's overall interactions. Generally, tests are included for: network and access-level testing, persistent data within the database cluster, authentication and authorization via the OAuth Server, session management via JWT tokens, web applications for different purposes, and log collection and storage policies. The methodology also includes considerations that may arise in specific projects implementing InterUSS, enabling the adaptability of the Testing Approach for applications across different targets.

As future work, we intend to validate the proposed methodology by applying it in production environments, addressing the operational complexities inherent in critical UTM infrastructures. Furthermore, we plan to empirically monitor emerging architectural modifications and custom implementations deployed by ecosystem authorities. By identifying these novel patterns, we aim to continuously investigate associated security threats, ensuring the testing guide remains a living document that evolves alongside the federated UAV traffic management landscape.

\section{Acknowledgment}

As part of the grammar and spelling check, Gemini was used on this work.

\bibliographystyle{sbc}
\bibliography{sbc-template}

@Misc{2021664,
  author       = {{European Commission}},
  title        = {{(EU) 2021/664}},
  year         = {2021},
}

@article{rv,
    author = {Asma Hamissi and Amine Dhraief and Layth Sliman},
    title = {A Review on Safety and Security in Decentralized Extensible Traffic Management Systems},
    journal = {SN Computer Science},
    year = {2025}
}

@Misc{monitoring,
  author       = {InterUSS Platform},
  howpublished = {https://github.com/interuss/monitoring},
  note         = {[Accessed at: May, 2026]},
  title        = {{InterUSS} monitoring v0.28.0},
  year         = {2026},
  url          = {https://github.com/interuss/monitoring},
}

@article{1,
    author = {Krishna Sampigethaya and Parimal Kopardekar},
    title = {Cyber security of unmanned aircraft system traffic management (UTM)},
    journal = {Integrated Communications, Navigation, Surveillance Conference (ICNS)},
    year = {2018}
}

@article{2,
    author = {Ralph Restituyo and Thaier Hayajneh},
    title = {Vulnerabilities and Attacks Analysis for Military and Commercial IoT Drones},
    journal = {IEEE Annual Ubiquitous Computing, Electronics \& Mobile Communication Conference},
    year = {2018}
}

@article{3,
    author = {Reham M. Fouda},
    title = {Security vulnerabilities of cyberphysical unmanned aircraft systems},
    journal = {IEEE Aerospace and Electronic Systems Magazine},
    year = {2018}
}

@article{4,
    author = {Rongxiao Guo and Buhong Wang and Jiang Weng},
    title = {Vulnerabilities and attacks of UAV cyber physical systems},
    journal = {International Conference on Computing, Networks and Internet of Things},
    year = {2020}
}

@article{5,
    author = {Chandra Sekar Veerappan and Peter Loh Kok Keong and Vivek Balachandran and Mohammad Shameel Bin Mohammad Fadilah},
    title = {DRAT: A Penetration Testing Framework for Drones},
    journal = {IEEE Conference on Industrial Electronics and Applications},
    year = {2021}
}

@article{6,
    author = {C.R.S. Kumar and Sanket Mohanty},
    title = {Current Trends in Cyber Security for Drones},
    journal = {International Carnahan Conference on Security Technology},
    year = {2021}
}

@article{7,
    author = {Jonas Gabrielsson and Joseph Bugeja and Bahtijar Vogel},
    title = {Hacking a Commercial Drone with Open-Source Software: Exploring Data Privacy Violations},
    journal = {Mediterranean Conference on Embedded Computing},
    year = {2021}
}

@article{8,
    author = {Fahad E. Salamh and Umit Karabiyik and Marcus K. Rogers and Eric T. Matson},
    title = {Unmanned Aerial Vehicle Kill Chain: Purple Teaming Tactics},
    journal = {IEEE Annual Computing and Communication Workshop and Conference},
    year = {2021}
}

@article{9,
    author = {Peng-Yong Kong},
    title = {A Survey of Cyberattack Countermeasures for Unmanned Aerial Vehicles},
    journal = {IEEE Access},
    year = {2021}
}

@article{10,
    author = {Gour Karmakar and Mark Petty and Hassan Ahmed and Rajkumar Das and Joarder Kamruzzaman},
    title = {Security of Internet of Things Devices: Ethical Hacking a Drone and its Mitigation Strategies},
    journal = {IEEE Asia-Pacific Conference on Computer Science and Data Engineering},
    year = {2022}
}

@article{11,
    author = {Jamison Colter and Matthew Kinnison and Alex Henderson and Stephen M. Schlager and Samuel Bryan and Katherine L. O’Grady and Ashlie Abballe and Steven Harbour},
    title = {Testing the Resiliency of Consumer Off-the-Shelf Drones to a Variety of Cyberattack Methods},
    journal = {IEEE/AIAA Digital Avionics Systems Conference},
    year = {2022}
}

@article{12,
    author = {Shihao Wu and Yang Li and Zhaoxuan Wang and Zheng Tan and Quan Pan},
    title = {A Highly Interpretable Framework for Generic Low-Cost UAV Attack Detection},
    journal = {IEEE Sensors Journal},
    year = {2023}
}

@article{13,
    author = {Réda Nouacer and Mahmoud Hussein and Paul Detterer and Eugenio Villar and Fernando Herrera and Carlo Tieri and Emmanuel Grolleau},
    title = {Towards a European Network of Enabling Technologies for Drones},
    journal = {DroneSE and RAPIDO: System Engineering for constrained embedded systems},
    year = {2023}
}

@article{14,
    author = {Aolin Ding and Matthew Chan and Amin Hass and Nils Ole Tippenhauer and Shiqing Ma and Saman Zonouz},
    title = {Get Your Cyber-Physical Tests Done! Data-Driven Vulnerability Assessment of Robotic Aerial Vehicles},
    journal = {Annual IEEE/IFIP International Conference on Dependable Systems and Networks},
    year = {2023}
}

@article{15,
    author = {Priyansh Sanghavi and Hargeet Kaur},
    title = {A Comprehensive Study on Cyber Security in Unmanned Aerial Vehicles},
    journal = {International Conference on Computing for Sustainable Global Development},
    year = {2023}
}

@article{16,
    author = {Ayusha Oli and Elmahedi Mahalal},
    title = {UAV Security: Attacks, Defenses, and Open Challenges},
    journal = {IEEE Access},
    year = {2025}
}

@article{17,
    author = {Jack Burbank and Toro Caleb and Emmanuela Andam and Naima Kaabouch},
    title = {Detection and Mitigation of Cyber Attacks on UAV Networks},
    journal = {MDPI Electronics},
    year = {2026}
}

@book{hacking_api,
    author = {Corey J. Ball},
    title = {Hacking APIs: Breaking Web Application Programming Interfaces},
    publisher = {No Starch Press},
    year = {2022}
}

@Misc{3411,
  author       = {{ASTM International}},
  title        = {{ASTM F3411-22 - Standard Specification for Remote ID and Tracking}},
  year         = {2022},
}

@Misc{3548,
  author       = {{ASTM International}},
  title        = {{ASTM F3548-21 - Standard Specification for UAS Traffic Management (UTM) UAS Service Supplier (USS) Interoperability}},
  year         = {2022},
}

@Misc{rfc_oauth_mtls,
  author       = {IETF},
  howpublished = {https://datatracker.ietf.org/doc/html/rfc8705},
  note         = {[Accessed at: May, 2026]},
  title        = {{RFC 8705 - OAuth 2.0 Mutual-TLS Client Authentication and Certificate-Bound Access Tokens}},
  year         = {2020},
  url          = {https://datatracker.ietf.org/doc/html/rfc8705},
}

@article{zero_trust,
    author = {Hongzhaoning Kang and Gang Liu and Quan Wang and Lei Meng and Jing Liu},
    title = {Theory and Application of Zero Trust Security: A Brief Survey},
    journal = {MDPI Entropy},
    year = {2023}
}

@Misc{wstg,
  author       = {OWASP},
  howpublished = {https://owasp.org/www-project-web-security-testing-guide/},
  note         = {[Accessed at: May, 2026]},
  title        = {{OWASP} Web Security Testing Guide v4.2},
  year         = {2020},
  url          = {https://owasp.org/www-project-web-security-testing-guide/},
}

@Misc{rfc_jwt,
  author       = {IETF},
  howpublished = {https://datatracker.ietf.org/doc/html/rfc7519},
  note         = {[Accessed at: May, 2026]},
  title        = {{RFC 7519 - JSON Web Token (JWT)}},
  year         = {2015},
  url          = {https://datatracker.ietf.org/doc/html/rfc7519},
}

@book{hackekshandbook,
    author = {Dafydd Stuttard and Marcus Pinto},
    title = {The Web Application Hackers Handbook 2nd Edition},
    publisher = {Wiley},
    year = {2011}
}

@article{oauth,
    author = {Daniel Fett and Ralf Küsters and Guido Schmitz},
    title = {A Comprehensive Formal Security Analysis of OAuth 2.0},
    journal = {2016 ACM SIGSAC Conference on Computer and Communications Security},
    year = {2016}
}

@article{literature_review,
    author = {Eda Marchetti and Tauheed Waheed and Antonello Calabrò},
    title = {Cybersecurity Testing in Drones Domain: A Systematic Literature Review},
    journal = {IEEE Access},
    year = {2024}
}

@article{raft,
    author = {Diego Ongaro and John Ousterhout},
    title = {In search of an understandable consensus algorithm},
    journal = {2014 USENIX Annual Technical Conference},
    year = {2014}
}

\end{document}